**Reversible Mode Switching in Y-coupled Terahertz Lasers**

Owen P. Marshall[1], Subhasish Chakraborty[1*], Md Khairuzzaman[1], Harvey E. Beere[2] and David A. Ritchie[2]

[1]School of Electrical and Electronic Engineering, University of Manchester, Manchester, M13 9PL, UK.

[2]Cavendish Laboratory, University of Cambridge, JJ Thomson Avenue, Cambridge, CB3 0HE, UK.

*Email: s.chakraborty@manchester.ac.uk

Electrically independent terahertz (THz) quantum cascade lasers (QCLs) are optically coupled in a Y configuration. Dual frequency, electronically switchable emission is achieved in one QCL using an aperiodic grating, designed using computer-generated hologram techniques, incorporated directly into the QCL waveguide by focussed ion beam milling. Multi-moded emission around 2.9 THz is inhibited, lasing instead occurring at switchable grating-selected frequencies of 2.88 and 2.92 THz. This photonic control and switching behaviour is selectively and reversibly transferred to the second, unmodified QCL via evanescent mode coupling, without the transfer of the inherent grating losses.

Burgeoning interest in terahertz (THz) spectroscopy and imaging is driving demand for highly functional THz sources with lower costs than those in widespread use today. Terahertz quantum cascade lasers (QCLs) are a leading contender for the role, offering milliwatts of coherent THz power from compact, electrically driven, semiconductor devices [1,2]. In the relatively young field of THz QCLs, many advances have been inspired by concepts developed in more mature, shorter wavelength semiconductor laser systems. Some of the most successful amongst these advances have been distributed feedback and photonic crystal lasers [3-6]. On the other hand, a number of ideas remain underdeveloped in THz QCLs. Notably, research is lacking into monolithic guided wave interconnects within these active THz systems. One of the simplest examples of such integrated systems is the Y-coupled laser. In the near- and mid-infrared regimes Y-coupled lasers have been employed for improved output powers [7], phase locking [8], non-mechanical beam shaping and steering [9,10], and numerous demonstrations of electronic frequency tuning [11-14]. Furthermore, the Y architecture naturally forms an elementary building block for more complex on-chip interferometric devices [15,16]. The potential to transfer this list of functionalities to THz QCLs makes Y-coupled structures highly attractive. Unfortunately, effects which rely upon driving current dependent variation in the effective refractive index ($n_{eff}$), such as many frequency tuning schemes, are problematic. Terahertz QCLs exhibit negligible electrical tuning of $n_{eff}$, and electronic tuning ranges are typically limited to a few GHz [17]. Nevertheless, the desire for electronically tunable single-mode THz QCLs motivated the recent development of an approach using a multi-band spectral filter embedded within a Fabry-Pérot (FP) cavity, with discrete tuning between filter resonances frequencies under the QCL gain bandwidth [18]. Without significant $n_{eff}$ tuning, the challenge is to create a multi-band spectral reflectivity $\rho(f)$, using a short structure compatible with practical THz QCL device sizes. By applying computer-generated hologram (CGH) design techniques based on Fourier transform principles to *longitudinal* structures, it is possible to create non-random yet aperiodic distributed feedback gratings with user defined responses in the frequency domain [18-20], as opposed to the spatial domain for a conventional *transverse* CGH [21]. The frequency resolution offered by the longitudinal CGH approach is inversely proportional to the hologram pixel count. Hence, relatively short aperiodic gratings generated in this way can possess multi-band ρ(*f*), with resonance frequency separations smaller than achievable with conventional photonic structures of equal length [19]. The holographic design approach not only offers control over individual resonance frequencies, but also their strengths. When a suitably designed aperiodic grating, (with reflectivity peak strengths limited to values matching the cleaved facets), is introduced to a longer FP QCL cavity, multi-mode emission is suppressed and single-mode lasing may be achieved on a single ρ(*f*) resonance. The precise resonance within which lasing occurs depends on the complex interplay of the grating, facet reflections and the QCL spectral gain. Crucially, within this compound grating-FP system, subtle changes to the spectral gain or complex $n_{eff}$ can alter the phase relationship between the aperiodic grating and cleaved facets [22]. This results in a highly dynamic mode competition landscape where single-mode switching can occur between ρ(*f*) resonances. In combination with coarse electrical gain tuning, this switching results in discrete single-mode electrical tuning of THz QCL emission [18,20].

In this letter the Y-coupling and longitudinal hologram concepts are combined in a THz QCL system. Firstly, two electrically independent, semi-insulating surface plasmon (SI-SP) THz QCL waveguides in a Y configuration are coupled via a shared evanescent substrate mode. An aperiodic grating is then introduced to just one of these QCLs to produce switchable single-mode emission. Finally, under appropriate electrical driving conditions this photonic control and frequency switching behaviour is selectively and reversibly transferred to the second, unmodified QCL. The transfer of spectral information takes place without the additional optical losses which would occur with direct incorporation of an aperiodic grating into the second QCL.

An outline of the basic Y structure and system behaviour is presented in Fig. 1. The schematic shows two independently electrically powered QCLs (*A* and *B*), which are purely optically coupled over a short section of their total length (shaded region). A multi-band ρ(*f*) is present in the arm of QCL *A*. Table 1(b) describes the influence of ρ(*f*) on the two arms under various powering combinations; emission from arm *A* is always dictated by ρ(*f*), whereas its control over arm *B* can be turned on and off. This system relies on the evanescent mode coupling which exists between SI-SP waveguides in close proximity on a shared substrate. Figure 2(a) shows a cross-sectional fundamental electromagnetic mode intensity profile for a single unperturbed SI-SP waveguide, calculated using the FIMMWAVE software package. The electromagnetic field has a large lobe extending well into the substrate. Figure 2(b) shows the mode intensity profile for a pair of neighbouring SI-SP waveguides with a ridge separation of 10 μm, revealing an overlap between the two substrate lobes. This is the target mode profile in the coupling region of Fig. 1(a).

Devices were fabricated from a molecular beam epitaxially-grown GaAs/Al$_{0.15}$Ga$_{0.85}$As active region (AR) based on reference 23. Previous straight ridge QCLs employing this AR and using standard FP cavities displayed multimode lasing around 2.9 THz, with a trend towards higher frequency modes at higher driving currents/biases due to variation in the 90 AR period lengths; shorter periods possessing higher laser transition energies and alignment biases [24]. Figure 2(c) shows the principal device dimensions. Each QCL is a 6.23 mm-long (the linear distance between, and perpendicular to, the parallel cleaved facets), 160 μm-wide SI-SP ridge waveguide. Contrary to many other Y-coupled semiconductor lasers in which waveguides were fully merged [7-14], the SI-SP QCLs share a substrate but their ARs remain physically distinct. They are separated by 10 μm over a length of ~0.5 mm, and then diverge via S-bends to a separation of 1 mm for the remaining ~3.3 mm device length. A bend radius of 3 mm was chosen to be as large as possible in order to minimise additional waveguide losses, while leaving sufficient length in the two straight waveguide regions for coupling and grating incorporation respectively. After heatsink packaging, QCLs were individually electrically contacted in an asymmetric two-terminal arrangement (found to have minimal impact on the performance of single straight ridges). Electrical independence of the QCLs was ensured by the selective removal of the highly-doped GaAs layer between the ridges by focussed ion beam (FIB) milling. The resulting 2 μm-wide by 2 μm-deep trench is visible in Fig. 2(c) running along the device line of symmetry. In a final post-processing step, (after fabrication and initial characterisation), an aperiodic grating was introduced to arm *A* as a series of narrow slots, FIB milled directly into the waveguide surface. Each slot was ~0.5 μm long and 100 μm wide (i.e. less than the ridge width) such that electrical contact was maintained with the entire length of QCL *A*, and only penetrated the upper metallic layers of the SI-SP waveguide, including the highly doped GaAs contact layer (top down: 20/180 nm Ti/Au, 100 nm PdGe, 80 nm $n^+$-GaAs doped at $5 \times 10^{18}$ cm$^{-3}$). Simulated cross-sectional mode intensity profiles, along with geometric considerations of the slot dimensions, lead to an estimated $n_{eff}$ perturbation for each slot of $|\Delta n_{eff}| = 0.1$. A symbolic representation of the grating, defined in terms of the minimum slot separation Λ, is given in Fig. 2(d). The total grating length is given by $L_{tot} = N\Lambda$, where the number of grating pixels is 2*N*. In this work the grating has *N* = 100 and Λ = 14.14 μm, giving $L_{tot}$ = 1.414 mm. There are 95 slots and ten Λ/2 phase shifts which act collectively to produce the target ρ(*f*) shown in Fig. 2(e), derived from the real-space dielectric constant distribution via their Fourier transform relationship [18-20]. It contains multiple resonances centred around a Nyquist frequency $f_N = c/2n_{eff}\Lambda$ = 2.88 THz, with a peak separation of ~58 GHz is given by $2f_N/N$. The target reflectivity values of the central three resonances around $f_N$ are ~30% (comparable to the cleaved FP facets), whereas outlying peaks are increased to ~40-50% in an attempt to extend the switchable emission range by compensating the reduced gain magnitude away from the gain centre frequency.

All testing was performed in pulsed operation at 4.5 K using synchronized pulse generators (1% duty cycle, 1 μs pulse length) and a Janis ST-100 continuous flow helium cryostat. Terahertz output powers were recorded without collection optics using a large area thermopile detector (calibrated to a Thomas Keating absolute THz power meter) placed inside the cryostat, directly in front of the 1 mm-separated QCL facets. Total system output powers (over ~2π solid angle) were measured in this arrangement, with no discrimination between individual facets. High resolution (0.075 cm$^{-1}$) emission spectra were recorded using a nitrogen-purged Bruker Vertex 80 Fourier transform infrared (FTIR) spectrometer and a helium-cooled bolometric detector.

Evidence of optical coupling between QCL arms was initially observed in the THz output power of the Y system. Voltage-current (V-I) and light-current (L-I) plots for QCLs *A* and *B* are presented in Figs. 3(a) and 3(b). When operated independently the performance characteristics are similar, with lasing threshold currents of 1.12 A and 1.11 A (~112 A cm$^{-2}$ and 111 A cm$^{-2}$), and peak pulsed output powers, $P_{max}$ of 4.6 mW and 4.9 mW (at ~159 A cm$^{-2}$ and 179 A cm$^{-2}$) respectively. Figure 3 also contains V-I and L-I data for each QCL when the other was held at two different fixed driving currents: just below, and well above lasing threshold. In the latter case we also subtract the initial lasing power offset due to the fixed-current QCL for ease of comparison (dotted lines). In all Y system L-I plots, the total output power is markedly higher than the linear sum of the maximum individual arm powers. For example, the maximum system power displayed in Fig. 3(b) is 9.7 mW, 15% higher than the combined individual $P_{max}$ recorded from arm *B* (4.9 mW) and the initial power offset from arm *A* (3.4 mW). Increased THz power is attributed to SI-SP waveguide coupling, enabling the AR gain of each arm to support and amplify the modal power of the other. Despite increased THz powers, the V-I characteristics remain unchanged, confirming good electrical isolation between QCLs. Note that similar power coupling effects were also witnessed before the aperiodic grating was introduced to arm *A* (not shown), and hence are an inherent property of the Y-coupled structure [25].

Prior to the introduction of the aperiodic grating, both Y arms displayed multi-moded emission and blue-shifting gain envelopes with increasing driving current. For example, Fig. 4(a) shows emission spectra from the unmodified arm *A* alone, across the entire operational driving current range of QCL *A*. Multi-moded behaviour was recorded at almost all driving currents, with frequencies primarily dictated by the round-trip phase solutions of the cleaved FP cavity. Note that the expected FP mode spacing of ~6.3 GHz for a 6.2 mm long QCL operating around 2.9 THz is modified by the presence of the waveguide S-bend which possesses its own dispersion characteristics. Many of the FP modes are suppressed and the remaining modes have unequal spacings of between ~6.3 and 19 GHz. A drastic redistribution of THz spectral power occurs after introduction of the grating to arm *A*. Figure 4(b) shows the observed switchable single-mode emission at frequencies of 2.88 and 2.92 THz. A side-mode suppression ratio of ~10 dB is observed upon switching to the higher frequency, though this could be improved by driving the laser at higher duty cycles. With the 1% duty cycle (chosen to minimise the device cooling requirements), the non-zero rise time of the driving current pulse always gives a small spurious contribution from lower driving currents, hence the persistent signal at 2.88 THz and the high-current 'tail' on the L-I plots in Fig 3. The two post-grating lasing modes are attributed to neighbouring reflectivity bands in Fig. 2(e). Although a one-to-one relationship between lasing modes and reflectivity bands is expected, modes will not necessarily occur at the same position within each band [18]. This can explain mode separations smaller than the 58 GHz band spacing. When a greater number of closely spaced bands are used, the average mode and band separations will tend to converge [18]. In contrast, when QCL *B* was operated singly the emission from the unmodified arm *B* remained multi-moded, even after the modification of arm *A*. Emission spectra collected from arm *B* with driving currents ranging from lasing threshold to ~$P_{max}$, are presented in Fig. 5(a). The output of arm *B* changes dramatically when both QCLs are operated simultaneously. Figure 5(b) shows the measured emission under almost identical conditions to Fig. 5(a), but with the addition of a constant driving current in QCL *A* ($I_A$ = 1.8 A), close to its individual $P_{max}$. At low QCL *B* driving current ($I_B$ = 1.1 A), only the small stray signal at 2.88 THz from arm *A* is detected. As $I_B$ is increased, prior multi-mode emission is suppressed, optical power instead appearing at the three grating-defined frequencies of 2.84, 2.88 and 2.92 THz. The strongest mode occurs at 2.88 THz, with a maximum modal power ~2.5 times that higher than the initial stray signal from arm *A*. Repeating this measurement with the collection optics focussed on the facet of arm *A*, a power rise of only 14% is recorded at 2.88 THz, confirming that the large increase originates from QCL *B*. We propose that evanescent mode coupling in the Y structure has enabled the high modal power in the grating-modified arm *A* to seed the unmodified arm *B*, thereby transferring user-defined spectral control. Furthermore, this control is reversible; multi-mode emission is re-established in the unmodified arm when the grating-modified QCL is turned off.

Transfer of spectral information between Y-coupled QCLs is not limited to a single resonance, but extends to the second grating defined spectral mode. Figure 6 shows emission spectra of QCLs *A* and *B*, operated singly and in combination with $I_A = I_B$ = 2.1 A. Individually, arm *A* predominantly lases upon the grating mode at 2.92 THz, arm *B* upon three modes at ~2.89, 2.91 and 2.93 THz. In combination, the unmodified QCL output is once again remotely controlled. The emission from arm *B* is largely constrained to 2.92 THz, evidenced by the increased power of this lasing mode and suppression of others. Hence by choosing appropriate driving currents in QCLs *A* and *B* it is possible to selectively establish frequency control and switching functionality in the latter, despite the absence of any directly imposed photonic structure within its waveguide.

In conclusion, optical coupling is achieved between two THz QCLs in a Y configuration on the same substrate. Frequency control and mode switching is accomplished in one QCL by introducing a holographically designed aperiodic grating. This behaviour is then transferred to the other electrically isolated QCL via purely optical coupling. The demonstration of coupling and switchable control in Y-coupled THz QCLs illustrates their potential as the basis for future systems, in which on-chip manipulation of spectral information is possible. This spectral control mechanism might be extended by coupling multiple control arms, each governed by a distinct aperiodic grating, to a single output laser. In this way, finely switchable emission over the entire QCL gain bandwidth could be achieved from a single source, removing any need for optical alignment to multiple lasers. Such a source would be beneficial in a number of spectroscopic applications, or in multi-wavelength THz imaging.

This work was supported by EPSRC First Grant EP/G064504/1 and partly supported by HMGCC.

FIG. 1. (a) Basic Y-coupled QCL schematic. A multi-band filter ρ(*f*) is introduced to just one of the two electrically independent, optically coupled QCLs using an aperiodic grating. Table (b) shows the controlling influence of ρ(*f*) over the QCL emission spectra under various electrical driving conditions.

FIG. 2. Simulated cross-sectional fundamental mode profiles of (a) a single ridge and (b) two closely spaced ridges. Blue through to yellow represent low to high mode intensities. (c) Photograph of Y-coupled THz QCLs, showing the electrical contacts, wire bonding, grating position, and critical device dimensions. (d) Symbolic representation of the aperiodic grating. Vertical lines represent milled slots with a minimum spacing Λ, dashes additional Λ/2 phase shift lengths. (e) Calculated grating spectral power reflectivity ρ(*f*). Multiple resonances occur within the estimated QCL gain bandwidth (shaded region).

FIG. 3. (a) V-I and L-I performance of QCL *A* (swept), with QCL *B* (fixed) under various driving conditions: unpowered (solid lines), just below lasing threshold (dashed), and well above threshold (dot dash). The dotted L-I corresponds to the latter condition after subtraction of the initial power offset from QCL *B*. (b) As above, with the lasers interchanged, i.e. *B* swept, *A* fixed. Arrows (i)-(vi) indicate the operating condition of the fixed-current QCL for the correspondingly labelled L-I of the swept laser.

FIG. 4. Emission spectra from arm *A* alone, across its entire operational current range, (a) before and (b) after FIB milling of the aperiodic grating. To aid comparison, spectral intensities are also projected vertically onto planar maps. Prior multi-mode emission is suppressed in favour of switchable single-mode emission. Insets: electrical powering and collection optics arrangements.

FIG. 5. (a) Emission spectra from arm *B* alone, from lasing threshold driving current to beyond $P_{max}$. (b) Measurements repeated with $I_A = 1.8$ A. Insets: electrical powering and collection optics arrangements.

FIG. 6. Emission spectra collected from the facet of arm *B*, with QCLs *A* (dotted line) and *B* (dashed) operated individually at driving currents of 2.1 A, and with both lasers operated simultaneously (solid). Inset: electrical powering and collection optics arrangement.

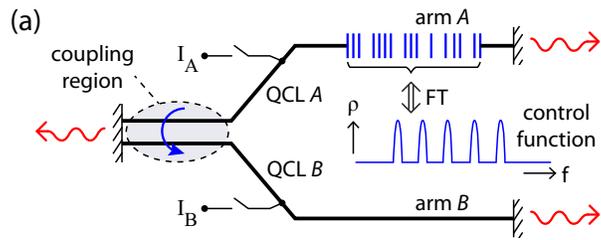

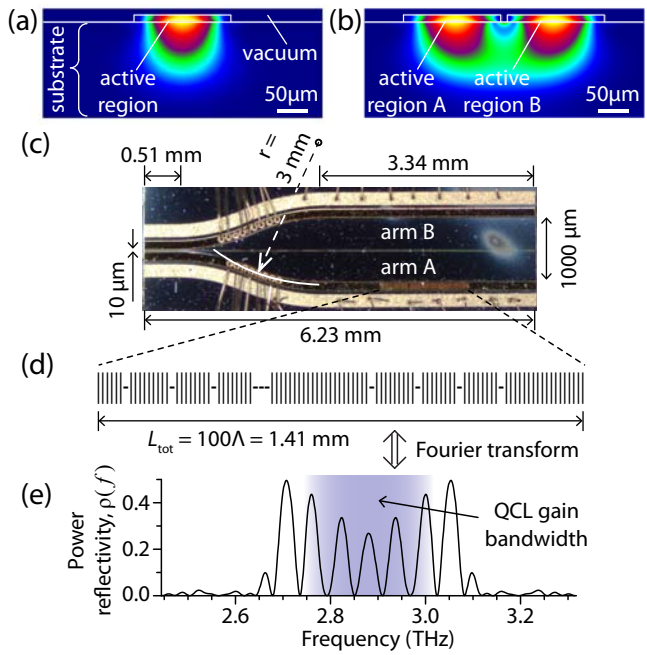

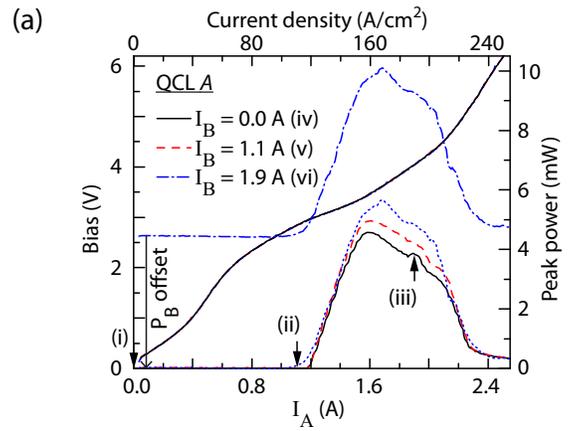

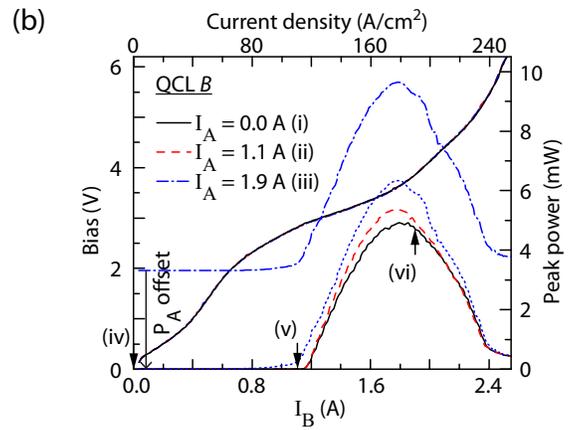

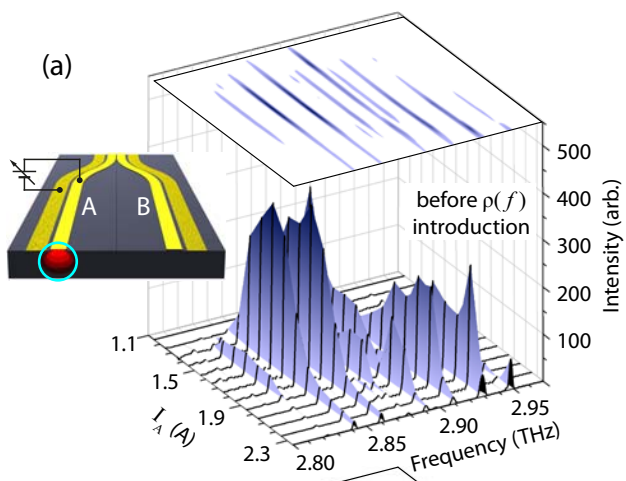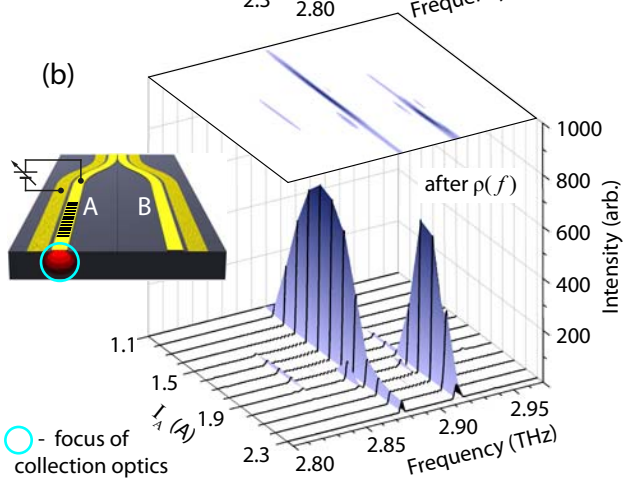

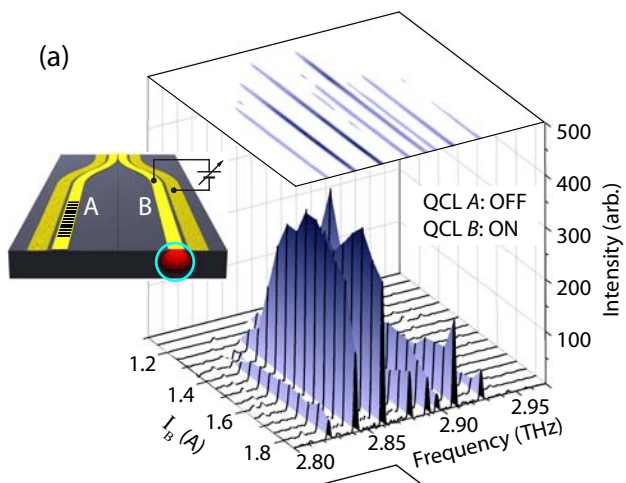
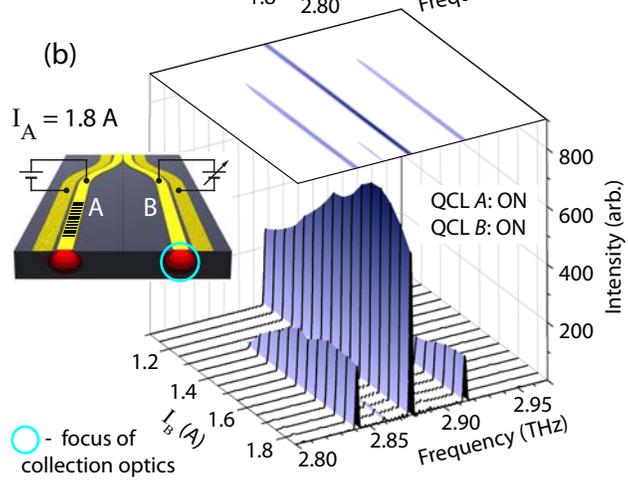

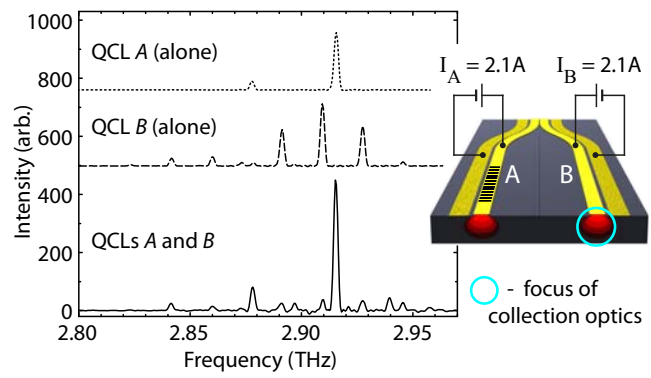